\documentclass[pra,floats,notitlepage,twocolumn]{revtex4-1}
\usepackage{latexsym}
\usepackage{amsmath}
\usepackage{amssymb}
\usepackage{bm}
\usepackage{wasysym}
\usepackage[dvips]{color}
\usepackage{graphicx}
\usepackage{graphicx}
\usepackage{subfigure}
\usepackage{epsfig}
\usepackage{pgfplots}
\usepackage{balance}
\DeclareMathAlphabet{\mathpzc}{OT1}{pzc}{m}{it}
\newcommand{\hide}[1]{}
\newcommand{\veps}{\varepsilon}

\def\bfp{{\bf p}}

\def\bfr{{\bf r}}
\def\ra{\rangle}
\def\la{\langle}

\def\veps{\varepsilon}
\def\ua{\uparrow}
\def\da{\downarrow}

\usepackage{tikz}
\usetikzlibrary{matrix}
\usetikzlibrary{arrows}

\begin{document}
\textheight=720pt
\title{Spin-orbit based devices for electron spin polarization}
\author{Y. Avishai$^{1,3,4}$}
\email{yshai@bgu.ac.il}
\author{Y. B. Band$^{2,3,}$}
\email{band@bgu.ac.il}
\affiliation{
${}^{1}$Department of Physics, and the Ilse Katz Center for Nano-Science, \\
${}^{2}$Department of Chemistry, Department of Physics, Department of Electro-Optics, and the Ilse Katz Center for Nano-Science, \\
Ben-Gurion University, Beer-Sheva 84105, Israel,\\
${}^{3}$New York University and the NYU-ECNU Institute of Physics at NYU Shanghai, 3663 Zhongshan Road North, Shanghai, 200062, China \\
$^{4}$ Yukawa Institute for Theoretical Physics, Kyoto University, Kyoto 606-8502, Japan}

\begin{abstract} 
We propose quantum devices having spin-orbit coupling (but no magnetic fields or magnetic materials) that, when attached to leads, yield a high degree of transmitted electron polarization.  An example of such a simple device is treated within a tight binding model composed of two 1D chains coupled by several consecutive rungs (i.e., a ladder) and subject to a gate voltage. The  ensuing scattering problem (with Rashba spin-orbit coupling) is solved, and a sizable polarization is predicted. When the ladder is twisted into a helix (as in DNA), the curvature energy augments the polarization.  For a system with random spin-orbit coupling, the distribution of polarization is broad, hence a high degree of polarization can be obtained in a measurement of a given disorder-realization. When disorder occurs in a double helix structure then, depending on scattering energy, the variance of the polarization distribution can increase even further due to helix curvature.
\end{abstract}
\maketitle

\section{Introduction.}
\label{Sec:_Intro}
Considerable interest has recently been focused on spintronic devices that give rise to spin polarization \cite{Sanvito_11, Urdampilleta_11, Charles_13}. Of particular importance for spintronics are devices that do not resort to external magnetic fields or magnetic materials \cite{Awschalom_09}, in contrast with those that do \cite{Upadhyay_99}.  A major advantage of semiconductor spintronic devices requiring only electric fields for manipulating spins is the lack of design complexities associated with incorporating local magnetic fields \cite{Awschalom_09}.  Interest in such devices was heightened following the measurement of spin-selective transmission of electrons through double-stranded DNA with spin polarizations exceeding 60\% at room temperature (the spin polarization efficiency depends on the length of the DNA and its organization) \cite{Gohler_11, Naaman_15}.  Here we propose a simple structure that yields outgoing polarized electrons given an incident beam of unpolarized electrons. It has a ladder-type structure (see Fig.~\ref{LadderE}) made of materials in which only spin-orbit (SO) is active (no magnetic fields). For simplicity and elegance, the SO is introduced as a non-Abelian gauge \cite{Frohlich}.

This paper is arranged as follows.  The description of the tight-binding model is presented in Sec.~\ref{Sec:_Model} and is used is Sec.~\ref{Sec:_Observables} to study electron polarization in a few mesoscopic systems: (1)  a clean planar and (2) a clean twisted two-chain ladder, in subsections \ref{subsec:_Plannar_Ladder} and \ref{Subsec:_Twisted_Ordered} respectively.  (3) In subsection~\ref{Subsec:_Twisted_Disoredered} we argue that a twisted disordered ladder emulates, in some sense, a DNA system and we evaluate the distribution of polarization for such system.   Our main results are summarized in Sec.~\ref{Sec:_Summary_Conclusion}, where we suggest the following: (a) For clean systems with the geometry of a planar ladder, one can achieve a high degree of electron polarization that is due solely to SO coupling. (b) The effect of twisting the ladder into double helical strands with links is reflected by a negative curvature energy that enhances the polarization dramatically when the Fermi energy is close to the band center. (c) For a twisted disordered system (reminiscent of DNA, in a sense to be discussed below), the distribution of polarization has zero mean, but it is broad enough to expect a sizable degree of polarization measured on a given disorder-realization. Moreover, the effect of curvature augments the variance of the distribution even further. Some relevant topics are clarified in the appendices, including the transfer matrix solution to the scattering problem, the definition of the Aharonov-Casher phase, a proof that in a system respecting time reversal symmetry with a single source and a single drain leads such that both leads are strictly one dimensional, there is no polarization, and a discussion of spin density, spin-current, spin-torque, and a useful relation between spin flux and spin-torque. 

\section{Tight binding Model}  \label{Sec:_Model}

\begin{figure} 
\centering
\centering{\includegraphics[width=0.45\textwidth]{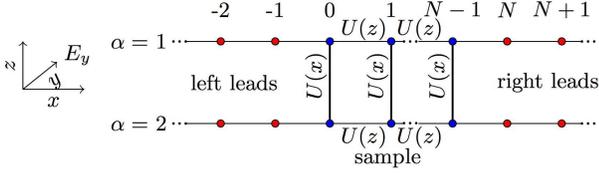}}
\caption{(Color online) Geometry of the device and the tight-binding model. Electrons hopping on two parallel chains (lying on the $x$-$z$ plane) can also hop between chains along $N$ adjacent consecutive rungs.  Free electrons moving in the leads (tight-binding sites marked by red points) are scattered off the sample in which SO is active (tight-binding sites marked by blue points).  Two cases are considered: (1) When the sample is acted upon by a homogeneous electric field $E \hat{\bf y}$, Rashba SO coupling generates SU(2) hopping matrices $U(z)=e^{\pm i \lambda \sigma_z}$ along horizontal links and $U(x)=e^{\pm  i \lambda \sigma_x}$ along vertical links, i.e., the rungs (this case is shown in the figure, and can be naturally extended to include other SO scattering mechanisms, e.g., Dresselhaus). (2) Any arbitrary specified inhomogeneous field configuration or Rashba/Dresselhaus coupling case can also be solved.  When the field configuration is too complicated for accurate description (e.g., in the case of DNA), the corresponding hopping can be encoded by random SU(2) matrices (in the same spirit as the treatment of energy levels of complex nuclei with random matrices).}
\label{LadderE}
\end{figure}

Guided by Fig.~\ref{LadderE}, we describe the system using a tight-binding model for an electron hopping on two chains numbered $\alpha = 1, 2$, where the sites on each chain are numbered $n=0, \pm1,\pm 2, \ldots$ and the lattice constant is $a$. Hopping between the two chains occurs along $N$ adjacent links. For $\alpha=1,2$, sites $n<0$ ($n>N-1$) form the left (right) leads, while sites $n = 0,1,...,N-1$ form the sample in which SO is active.  Hopping between two sites $(\alpha,n) \leftrightarrow (\alpha,n\pm 1)$ and  $(\alpha,n) \leftrightarrow (\beta,n)$ {\it within the sample} is encoded by SU(2) matrices $e^{i \lambda \hat {\bf n}_m \cdot {\bm \sigma}}$, where $\lambda$ is the dimensionless spin-orbit strength parameter, ${\bm \sigma}$ is the vector of Pauli spin matrices, and the unit vector $\hat {\bf n}_m$ determines the direction of the effective magnetic field acting on the electron on link $m$ (either along a chain or between chains).  The representation of SO coupling in terms of SU(2) matrices, an extension of the Peierls substitution \cite{Spohn} for a non-Abelian gauge, is discussed elsewhere, see e.g., Refs.~\cite{Ando, OEW}.  In semiconductors there are various sources of SO coupling\cite{RH}, such as the Dresselhaus term (spin component parallel to the wave number ${\bf k}$) and the Rashba term due to structural inversion asymmetry (spin component perpendicular to ${\bf k}$), or any combination thereof. The SO interaction strength is significantly enhanced compared to its value in vacuum.  Moreover, the SO strength and Fermi energy can be controlled by applying gate voltages (see Ref.~\cite{Nitta}, which studies an array of mesoscopic InGaAs rings controlled by Au gate electrodes).

The model Hamiltonian for the device in Fig.~\ref{LadderE} is written in second-quantization using the annihilation operator $\hat {c}_{\alpha n \sigma}$ for an electron at site $(\alpha,n)$ with spin projection $\sigma = \, \ua, \da$, 
\begin{eqnarray}
H &=&  -t \sum_\alpha \! \! \left[\sum_{n=0}^{N-2} \hat{c}^\dagger_{\alpha  n}  e^{ i \lambda \hat {\bf n}_n \cdot {\boldsymbol \sigma}} \hat{c}_{\alpha n+1} \! + \!\! \sum_{n \notin[0,N-1]} \hat{c}^\dagger_{\alpha  n} \hat{c}_{\alpha n+1}\right] 
  \nonumber \\
&&-t\sum_{n=0}^{N-1} \hat{c}^\dagger_{1  n} \, e^{ i \lambda \hat {\bf n}_n \cdot {\boldsymbol \sigma}} \, \hat{c}_{2  n}  + \mathrm{h.c.},
   \label{H}
\end{eqnarray}
where $\hat {c}_{\alpha n}$ =$ (\hat {c}_{\alpha n \ua},\hat {c}_{\alpha n \da})^T$ and $t$ is the (real) hopping amplitude (we take $t=1$  in what follows).

\subsection{The scattering Problem} \label{Subsec:_Scatt}
The Schr\"odinger equation with scattering boundary conditions is solved by the transfer matrix technique, see Appendix \ref{t_matrix}.  For definiteness, we consider scattering of an incoming electron at Fermi energy $\varepsilon = -2  \cos ka$ as it approaches the sample from the left in channel $\beta=1,2$ with spin direction $\mu=\ua,\da$. It can be reflected or transmitted into channel $\alpha = 1,2$ with spin direction $\sigma=\ua,\da$.  The reflection and transmission amplitudes are the $4$$\times$$4$ matrices $r_{ \alpha \sigma \beta \mu}$ and $t_{ \alpha \sigma \beta \mu}$, 
where the initial and final states are denoted by $| \beta \mu \ra$ and $|\alpha \sigma \ra$ respectively. 
Unitarity and time-reversal constraints imply \cite{band-avishai-1},
\begin{eqnarray} \label{unitarity}
&&{\mathrm{Tr}}[t^\dagger t + r^\dagger r] = 4, \nonumber \\
&& t'_{\alpha \nu \beta \mu} = (-1)^{\nu-\mu}t^*_{\beta \bar{\mu }\alpha \bar{\nu}}, \ r_{\alpha \nu \beta \mu} = (-1)^{\nu-\mu}r_{\beta \bar{\mu } \alpha \bar{\nu}},
\end{eqnarray}
where $t'$ is the transmission matrix for scattering of incoming electrons from the right, and $\bar {\sigma} = -\sigma$.  
Time reversal invariance implies, $r_{\alpha \sigma \alpha \mu} = \rho \, \delta_{\nu \mu}$ where $|\rho| \le 1$.
\subsection{Order of magnitude of spin order strength $\lambda$}
\label{Subsec:_lambda}
Let $\alpha_R$ denote the SO strength parameter defined through the Rashba Hamiltonian, $H_{\mathrm{R}}=\frac{\alpha_R}{\hbar} (\bfp \times {\bm \sigma})\cdot \hat {\bf z}$, for a two-dimensional electron gas in an asymmetric quantum well, and let the lattice constant $a \approx 1 \, \mu$. The dimensionless spin-orbit strength parameter is $\lambda = m^*a \alpha_R/\hbar^2$ where $m^*$ is the effective electron (or hole) mass\cite{Cardona}.  $\lambda$ can be varied in the range, $0 \le \lambda \le 4$ (e.g., see Ref.~\cite{Nitta} which studied SO effects on the conductance of a 2D network of InGaAs rings with radius $r \approx 1 \, \mu$).
\section{Observables}
\label{Sec:_Observables}
We are mainly interested in the dimensionless conductance\cite{Landauer}$g$, and the polarization of the transmitted electrons, $P_z  =(N_\ua -N_\da)/(N_\ua + N_\da)$, where $N_\sigma$ is the number of transmitted electrons with spin projection $\sigma = \ua,\da$.
Generically, the transmitted and reflected polarizations ${\bf P}_T$ and ${\bf P}_R$ are vectors in spin space. Thus,
\begin{equation} \label{BAPol}
g= {\mathrm{Tr}} [t^\dagger t], \quad {\bf P}_T = \frac{\mathrm{Tr}[t^\dagger {\bm \Sigma} t]}{g} , \quad {\bf P}_R = \frac{\mathrm{Tr}[r^\dagger {\bm \Sigma} r]}{g} ,  
\end{equation}
where ${\boldsymbol \Sigma}= I_{2 \times 2} \otimes {\bm \sigma}$.  All physical measurable observables (see Appendix \ref{App_spin-torque}), such as $g$ and ${\bf P}$, depend on the spin-orbit strength $\lambda$, the wave number $k$ (equivalently on the Fermi energy $\veps =-2 \cos ka$) and the number of rungs $N$ (equivalently the length $L$ of the sample). We shall see below that the behavior of these quantities as functions of $k$ for fixed $\lambda$ or vice versa, are qualitatively similar because $\lambda$ appears together with $k$ in the expression for the covariant wave number $(k-\frac{\lambda}{a} \, \hat{\bf n} \cdot {\bm \sigma})$. This is completely analogous to the $U(1)$ case where the vector potential (e.g., for an electron in a ring) is present in the covariant wave number $(k-\phi_{\mathrm{AB}}/a)$, where $\phi_{\mathrm{AB}}$ is the dimensionless magnetic flux through the ring of radius $a$, responsible for the Aharonov-Bohm effect.  For a single square in our ladder, $\lambda$ is directly related to the Aharonov-Casher phase, $\lambda_{\mathrm{AC}}$ (see Appendix \ref{AC_phase}). Note that $\phi_{\mathrm{AB}}$ is directly proportional to the strength of the U(1) vector potential, but the relation between $\lambda_{\mathrm{AC}}$ and the coupling strength $\lambda$ of the SU(2) vector potential is less simple, 
\begin{equation} \label{phiAC}
\cos \lambda_{\mathrm{AC}} =1-2 \sin^4 \lambda.
\end{equation}
\subsection{Plannar ladder} \label{subsec:_Plannar_Ladder}
\begin{figure}
\centering
\centering\subfigure[]{\includegraphics[width=0.40\textwidth]{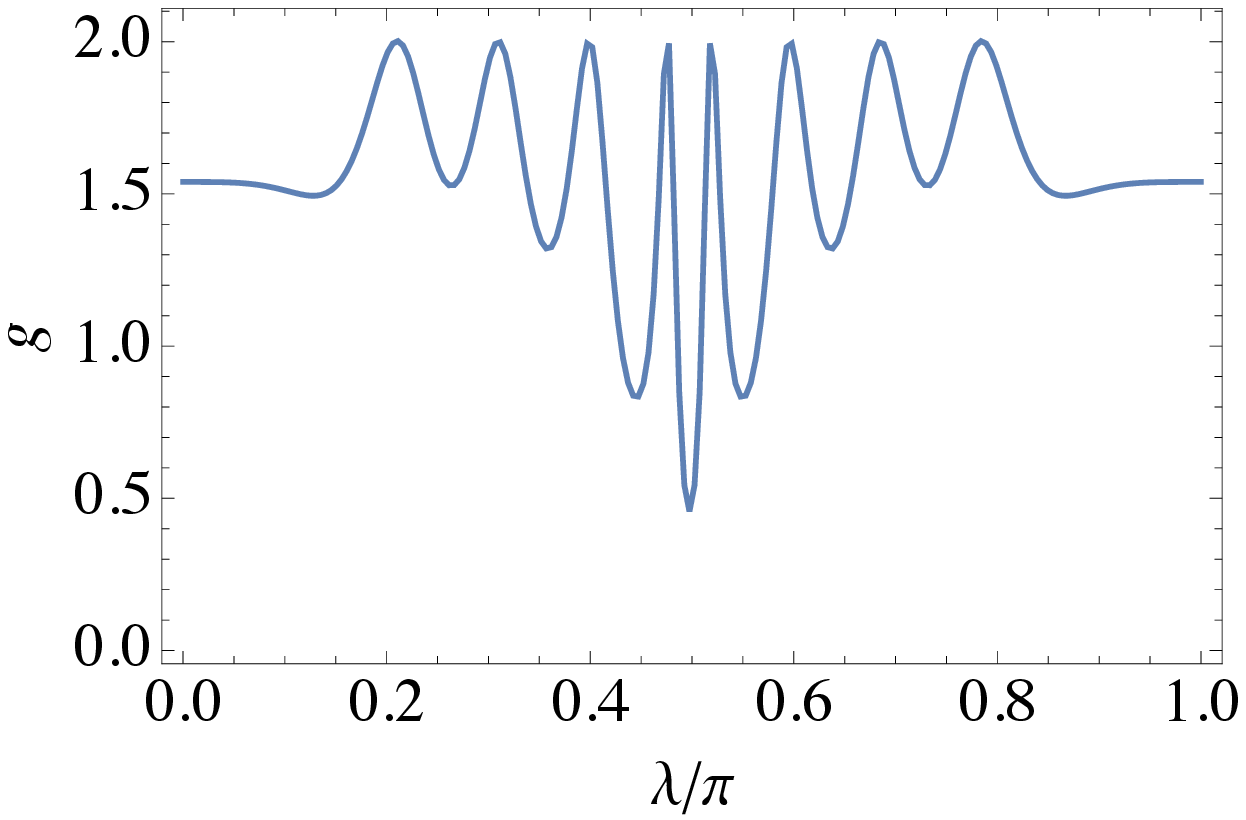}}
\centering\subfigure[]{\includegraphics[width=0.40\textwidth]{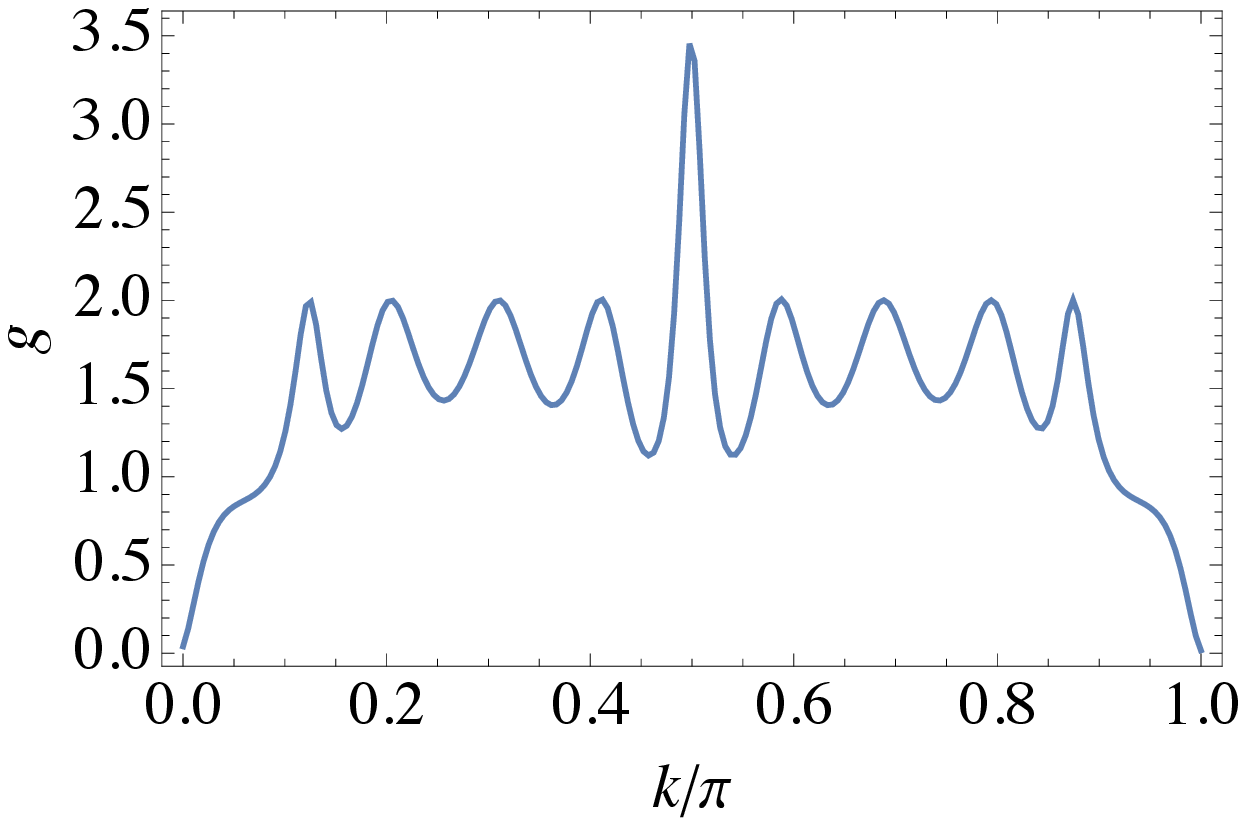}}
\centering\subfigure[]{\includegraphics[width=0.40\textwidth]{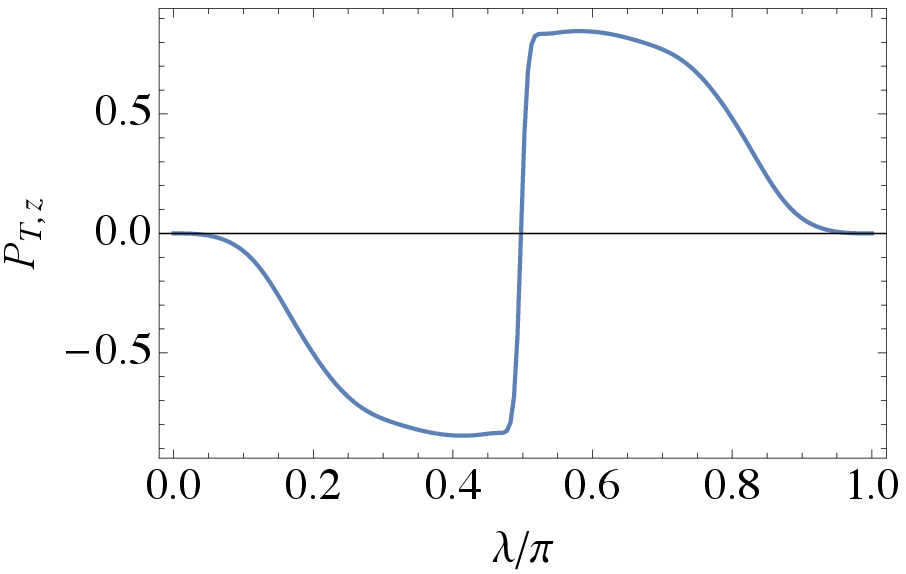}}
\centering\subfigure[]{\includegraphics[width=0.40\textwidth]{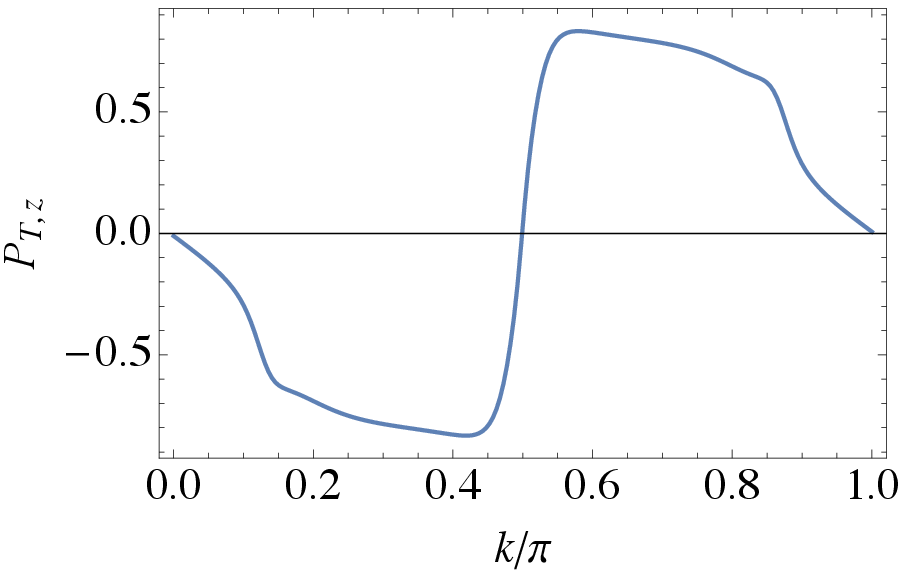}}
\caption{Results for 9 links: (a) Conductance as function of $\lambda$ with $k=1$. (b) Conductance as function of $k$ with $\lambda=1$.  (c) $P_{T,z}$ as function of $\lambda$ with $k=1$. The sharp variation, accompanied by sign change, around $\lambda= \ \pi/2$ is discussed in the text. (d) $P_{T,z}$ as function of $k$ with $\lambda=1$. The results shown in (c) and (d) imply that the sign of polarization can be controlled either by tuning the Fermi energy or by tuning the strength of the SO coupling.}
\label{FigLadder1}
\end{figure}

The strength of the Rashba SO coupling due to structural inversion asymmetry can be controlled by applying a gate voltage generating a perpendicular homogeneous electric field ${\bf E} \parallel \hat {\bf y}$ on sites $n \in [0,N-1]$, see Fig.~\ref{LadderE}. The corresponding SU(2) matrices in  Fig.~\ref{LadderE} are then, $U(x)=e^{\pm i \lambda \sigma_x}$ and $U(z)=e^{\pm i \lambda \sigma_z}$.  In Fig.~\ref{FigLadder1} we plot the conductance $g$ and the transmitted polarization $P_{T,z}$ for $N=9$ links.  In (a) $g$ is plotted for fixed $k$ as a function of $\lambda$ while in (b) $g$ is plotted for fixed $\lambda$ as function of $k$. The pattern of the conductance is characterized by a series of peaks reflecting the mini-band structure commensurate with the number of links. As anticipated in our discussion above, the qualitative similarity of the two patterns is evident, except at the band center where resonant transmission sometimes occurs, and at the band edges $k = 0 , \pi$ where the conductance vanishes.  The corresponding transmitted polarizations are shown in (c) and (d). At the peaks, the magnitude of the polarization is remarkably high. Achieving 80\% polarization with such a simple structure based solely on spin-orbit coupling is a godsend for spintronics.  Just like the patterns of conductance in (a) and (b), the similarity between the  patterns of polarization shown in (c) and (d) is evident.

As is evident from Figs.~\ref{FigLadder1}(c) and \ref{FigLadder1}(d), $P_{T,z}(k,\lambda)$ varies sharply and changes sign at $k=\pi/2$ (the band center)  and at $\lambda = \pi/2$.  Consequently, the sign of the polarization is controllable by a gate voltage, an attractive feature from the spintronics point of view.
The sharp slope of $P_{T,z}$ near the sign reversal point $\lambda = \pi/2$ occurs due to the term $(i \sin \lambda \, \hat {\bf n} \cdot {\bm \sigma})$ [that controls the spin physics within the hopping matrix $e^{i \lambda \hat {\bf n} \cdot {\bm \sigma}}= \cos \lambda+i \sin \lambda \, \hat {\bf n} \cdot {\bm \sigma}$] which is maximal at $\pi/2$. 

\subsection{Twisted ordered ladder}  \label{Subsec:_Twisted_Ordered}
Now we modify the geometry to be closer to that of DNA by twisting the ladder so that each strand becomes a helix\cite{Matityahu}.
After writing the Schr\"odinger equation for an electron on a single 1D helical strand, we treat the dynamics of an electron on the twisted ladder within a tight-binding model.  The resulting  Hamiltonian is similar to that of Eq.~(\ref{H}), albeit with one modification: The kinetic energy term includes a constant negative energy due to the curvature of the helix. The consequences of this modification will be analyzed for ordered and disordered ladders below.  We start from Eq.~(9) of Ref.~\cite{Gutierrez} and modify it such that the SO enters as an SU(2) gauge.  Using $s$ as a length coordinate along the helix, the Hamiltonian for an electron moving along a (strictly) 1D helix is, 
\begin{equation} \label{G3}
H_{1D} = \frac{\hbar^2}{2 m} \left (-i \frac{\partial}{\partial s}+\frac{e}{\hbar c}{\cal A} \right )^2 - \frac{\hbar^2 \kappa^2}{8m}+V(s) ,
\end{equation}
where ${\cal A} = \tfrac{\hbar}{4 mc} {\bm \sigma} \times {\bf E}(s)$ is a local space-dependent SU(2) vector potential \cite{Gutierrez}.  Here $V(s)$ is a local potential (assumed to be 0 for simplicity), and $\veps_c = -\tfrac{\hbar^2 \kappa^2}{8m}$ is the curvature energy, where $\kappa$ is the curvature of the helix.

Before turning to the tight-binding formulation, it is important to point out that one should not expect polarization in a strictly 1D structure because, without a closed loop, the vector potential can be eliminated by a suitable gauge transformation (see Appendix \ref{Sed:NoP}). Moreover, in 1D, time reversal invariance, Eq.~(\ref{unitarity}), implies that the 2$\times$2 reflection matrix $r \propto 1_{2\times 2}$ so that according to Eq.~(\ref{BAPol}), ${\bf P}_R = 0$. Similar considerations hold for the transmitted polarization, hence ${\bf P}_T=0$.

For the ordered ladder, we assume that the electric field ${\bf E}$ is also ``twisted'' in the sense that its action on a given link in the twisted and non-twisted ladders is the same.  Consequently, the sole effect of twisting is the occurrence of the curvature energy. If the length coordinate $s$ is expressed in units of $(1/\kappa)$ then energies are measured in units of $\tfrac{\hbar^2 \kappa^2}{2 m}$ and $\veps_c$ = -1/4 (dimensionless). Thus, the energy on sites $(\alpha,n)$, $\alpha = 1,2$, and $n = 0,1,2,\ldots,N-1$, is lowered by $\veps_c$.  The effect of curvature is expected to be significant when $|\veps_c| > |\veps|=|2 \cos k a|$.  Hence, the wave number is chosen close to the band center where the Fermi energy is small.

\begin{figure} 
\centering
\centering\subfigure[]{\includegraphics[width=0.45\textwidth]{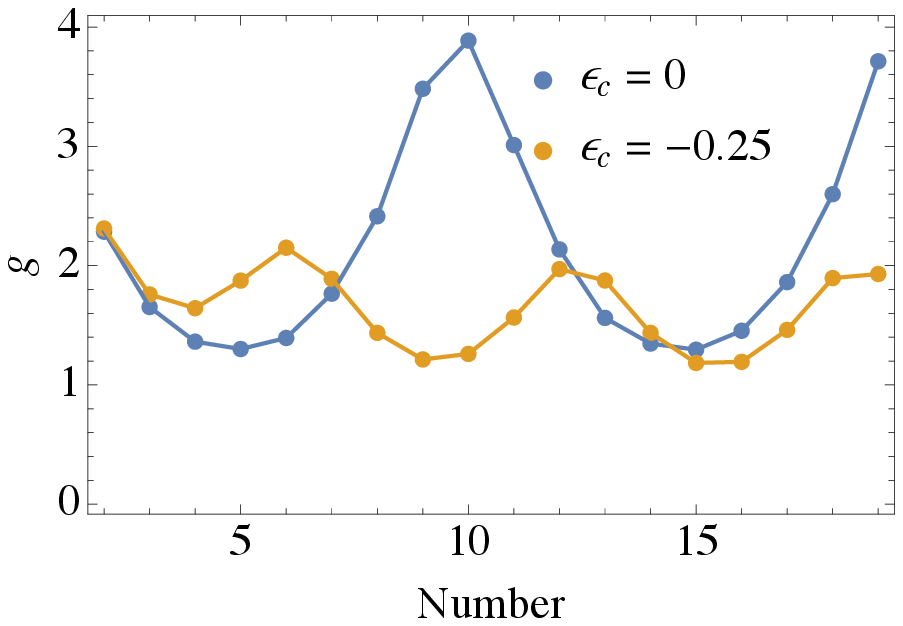}}
\centering\subfigure[]{\includegraphics[width=0.45\textwidth]{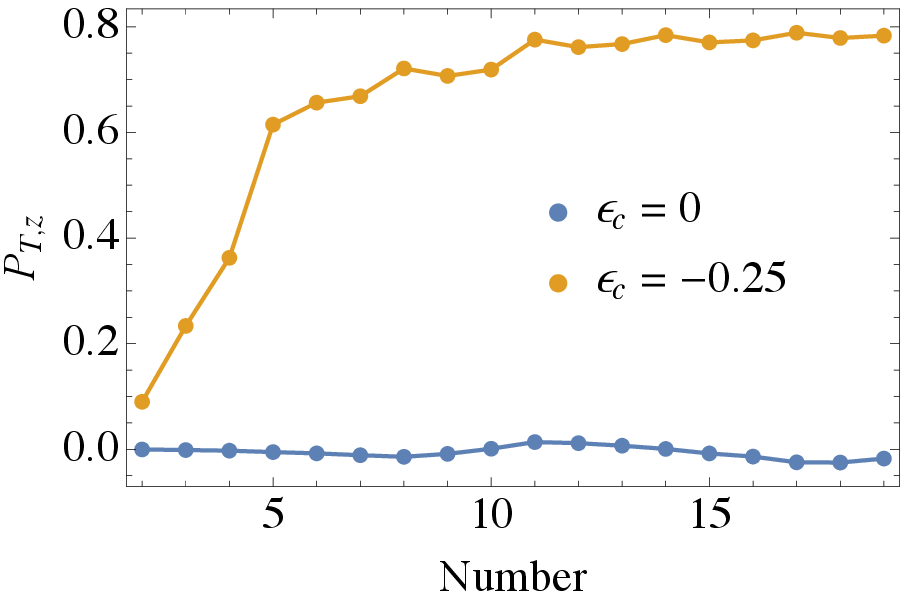}}
\caption{(Color online) (a) Conductance $g$ and (b) polarization $P_{T,z}$ as function of number of links $N$ with $k= 1.57$ and $\lambda=1$ for the non-twisted ladder (as shown in Fig.~\ref{LadderE}, with zero curvature $\veps_c=0$) and for the twisted ladder  with finite curvature   $\veps_c=-1/4$.}
\label{FigLadder2}
\end{figure}

The conductance $g$ and the polarization $P_{T,z}$ are plotted in Fig.~\ref{FigLadder2} versus $N=2$, 3, $\ldots, 19$ (the number of links) for $k=1.57$ and $\lambda=1$ for the non-twisted ladder ($\veps_c=0$, as in Fig.~\ref{LadderE}) and for the twisted ladder ($\veps_c = -1/4$).  The curvature has a dramatic effect on the polarization. In the absence of curvature, the polarization at the band center virtually vanishes, but when the curvature is included, the polarization saturates close 80\%. This result depends on the position of the Fermi energy; far away from the band center, the polarization generally remains high, but the curvature has just a small effect. 

\subsection{Combination of twisting and disorder}  \label{Subsec:_Twisted_Disoredered}
We now turn to the case where the SO coupling on the various links are not solely determined by the Rashba SO mechanism, and where the underlying electric fields are not necessarily homogeneous.  For example, in DNA \cite{Gohler_11, Naaman_15}, the pattern of local fields responsible for SO coupling (strengths, directions, etc.) is extremely complicated and may vary sharply as function of position.  We model this by assuming that the Hamiltonian (\ref{H}) is a random matrix drawn from a GSE. 
Similar concepts are prevalant in many branches of physics since the early days of random-matrix theory that was employed for analyzing the energy spectra of complex nuclei.  To implement this in our disordered ladder, the SU(2) hopping matrices on the links are assumed to be random, independently and identically distributed.  
Now, the problem of generating an ensemble of these matrices must be addressed.  An SU(2) matrix is determined by three Euler angles,
\begin{equation} \label{SU2Euler}
U = \begin{pmatrix} e^{i \gamma} \cos \theta & e^{i \beta} \sin{\theta}\\-e^{-i \beta} \sin \theta & e^{-i \gamma} \cos \theta \end{pmatrix},
\end{equation} 
where $0 \le \theta \le \pi$, $0 \le \beta \le 2 \pi$, $0 \le \gamma \le 2 \pi$.  Generating random SU(2) matrices should take into account the Haar measure, $d\mu(\gamma, \beta,x)=\frac{1}{2} d[x^2] d \beta d \gamma=x  d \beta d \gamma dx$, where $x = \cos \theta$. We follow the method introduced in Ref.~\cite{Ozols}, but with a trivial modification \cite{Ozols_mod}.  

The calculation of the distribution is now carried out employing a large ensemble of random SU(2) matrices.  Our aim here is to elucidate the distribution $D(P_{T,z})$ of the transmitted polarization and its two lowest moments for a given number of links $N$ and wave number $k$. The transmitted polarization is expected to be finite in all three directions in spin space, i.e., $P_{T,x}$, $P_{T,y}$ and $P_{T,y} \ne 0$, but here we focus for simplicity on $P_{T,z}$. It is expected that $\la P_{T,z} \ra = 0$, but the interesting quantity here is the variance $v(P_{T,z}) \equiv \la P_{T,z}^2 \ra$ because large variance means that there is a significant probability for detecting a high degree of polarization {\em in a measurement of a given disorder-realization}.

\begin{figure}
\centering
\centering{\includegraphics[width=0.45\textwidth]{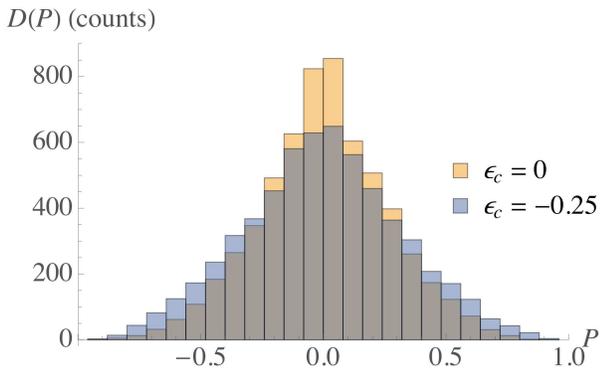}}
\caption{(Color online)
Polarization distributions $D(P_{T,z})$ in a disordered ladder with 15 links, for fixed wave number $k=1.5$, based on ensembles of 6000 samples. 
A comparison is made between the cases of $\veps_c=0$ (orange) and $\veps_c=-1/4$ (blue), see Eq. (\ref{G3}), (corresponding to non-twisted and twisted ladder respectively). The first moment vanishes for both distributions but the variances (and also the average conductance, not shown here) are very different; for $\veps_c=0$ we find $\la g \ra=0.211$, $v(P_{T,z})=0.0689$ while for $\veps_c=-1/4$ we find $\la g \ra=0.629$, $v(P_{T,z})=0.101$.  Thus, the effect of curvature is to augment the conductance by a factor $\simeq 3$ and the variance of the polarization distribution by a factor $\simeq 1.5$.
}
\label{FigLadder3}
\end{figure}

Fig.~\ref{FigLadder3} shows the distributions $D(P_{T,z})$ for $\veps_c = 0$ and $\veps_c = -1/4$ for a ladder of 15 links with $k=1.5$.  {\em The effect of curvature  is  to enhance the variance of the polarization distribution} (see the figure caption for details). This augments the probability to achieve higher polarization in a measurement of a given disorder-realization. Thus, by tuning the Fermi energy close to the band center we can substantially increase the width of the polarization distribution.
\section{Summary and conclusions} \label{Sec:_Summary_Conclusion} 
We have shown that SO coupling in a 2D mesoscopic device having a simple ladder geometry yields a high degree of transmitted electron polarization. Moreover, similarly to the study of conductance in InGaAs samples \cite{Nitta}, applying a gate voltage allows control of the polarization either by tuning the SO strength or the electron density (equivalently, the Fermi energy).  When the ladder is twisted into a helix (as in DNA), the curvature energy plays an important role by increasing the polarization, particularly for Fermi energy near the band center.  For complicated systems with rapidly and randomly varying local electric fields, it is reasonable to assume that the parameters determining the local SO strength $\lambda$ and direction $\hat {\bf n}$ are independently and identically distributed random numbers, i.e., the hopping terms $e^{i \lambda \hat{\bf n} \cdot {\bm \sigma}}$  appearing bin Eq.~(\ref{H}) are random SU(2) matrices. The resulting distribution of transmitted polarization is broad enough to expect a high degree of polarization in a measurement of a given disorder-realization. The variance of the distribution can be increased even further, as shown in Fig.~\ref{FigLadder3}, due to the curvature of the helix.  The results reported here show that simple spintronic devices that polarize electrons can be designed without resorting to the use of magnetic materials or external magnetic fields\cite{Hatano}.   

Finally, it is worth mentioning the relevance of Refs.~\cite{Miron} and \cite{Liu} to our work. Reference~\cite{Miron} demonstrated that a perpendicularly magnetized cobalt quantum dot can be switched at room temperature by injecting an in-plane current.  The symmetry of the switching field is consistent with the spin accumulation induced by Rashba interaction and also with the torque induced by the spin Hall effect.  The relevance to our work is that the effective magnetic field that drives the switching is due to Rashba SO coupling.  However, our motivation is entirely different, as we are interested in obtaining a current of polarized electrons and not in reversing the polarization of a magnetic material.  Reference~\cite{Liu} uses the spin Hall effect to generate a strong spin current that induces efficient spin-torque switching of ferromagnets at room temperature. Here also, our motivation is very different, as explained above.  In both these works, the relevance of spin-torque is stressed. Indeed, spin-torque is an important ingredient in contemporary spintronics.  Using the continuity equation for spin currents, we show in Appendix \ref{App_spin-torque} that the volume integrated spin-torque of the ladder system is directly related to the transmitted and reflected polarizations defined in Eq.~(\ref{BAPol}).  

\bigskip

\noindent
{\it Acknowledgement.} This work was supported in part by grants from the Israel Science Foundation:  No.~400/2012 (YA), No.~295/2011 (YBB),  and  DFG Grant FO 703/2-1 through the DIP program  (YBB).  YA  thanks D. Ariad, C. de Morais Smith and Z. Ovadyahu for fruitful discussions.
\appendix
\section{Solution of the ladder model by the Transfer Matrix Method} \label{t_matrix}

Here we solve the scattering problem for the model whose Hamiltonian is introduced in Eq.~(1) (re-written here for self-consistency) and compute transmission and reflection amplitudes.  The model Hamiltonian for the device in Fig.~1 is written in second-quantization using the annihilation operator $\hat {c}_{\alpha n \sigma}$ for an electron at site $(\alpha,n)$ with spin projection $\sigma = \, \ua, \da$, 
\begin{eqnarray}   \label{H}
&&H =  -t \sum_\alpha \! \! \left[\sum_{n=0}^{N-2} \hat{c}^\dagger_{\alpha  n}  e^{ i \lambda \hat {\bf n}_n \cdot {\boldsymbol \sigma}} \hat{c}_{\alpha n+1} \! + \!\! \sum_{n \notin[0,N-1]} \hat{c}^\dagger_{\alpha  n} \hat{c}_{\alpha n+1}\right] 
\nonumber \\
&& -t\sum_{n=0}^{N-1} \hat{c}^\dagger_{1  n} \, e^{ i \lambda \hat {\bf n}_n \cdot {\boldsymbol \sigma}} \, \hat{c}_{2  n}  + \mathrm{h.c.},
\end{eqnarray}
where $\hat {c}_{\alpha n}$ =$ (\hat {c}_{\alpha n \ua},\hat {c}_{\alpha n \da})^T$ and $t$ is the (real) hopping amplitude.

Our aim is to solve the Schr\"odinger equation $H |\Psi \ra = \veps |\Psi \ra$ for the two component spinor $|\Psi \ra$, subject to 
scattering boundary conditions.  Here $\varepsilon=-2 \cos k$ is the scattering energy and $k$ is the wave number (where we have taken the lattice constant $a=1$).  For definiteness, we consider a scattering problem wherein an incoming electron approaches the link at $n=0$ from the left ($n<0$) in channel $\beta=1,2$ with spin direction $\mu=\pm=\ua,\da$. It can be reflected or transmitted into channel $\alpha = 1,2$ with spin direction $\sigma=\pm=\ua,\da$.  Henceforth, the spinor wave functions and the scattering amplitudes depend on (and should carry) the initial quantum numbers $|\beta \mu \ra$. Thus, the corresponding reflection and transmission amplitudes are  written as $r_{ \alpha \sigma;\beta \mu}$, and $t_{ \alpha \sigma;\beta \mu}$. 

We expand the spinor in a complete set of basis functions in the [chain$\otimes$site$\otimes$spin]  space.  The basis functions are denoted by $|\alpha n  \sigma \ra$; explicitly, $|\alpha  n \ua \ra=|\alpha n \ra \otimes \binom{1}{0}$ and $|\alpha  n \da \ra = |\alpha n \ra \otimes \binom{0}{1}$. Thus,  
\begin{equation} \label{Psi}
| \Psi \ra_{\beta \mu} = \sum_{\alpha n \sigma } \psi_{\alpha \sigma;\beta \mu}(n) |\alpha n \sigma \ra~, \ \ \psi_{\alpha;\beta \mu}(n) = \begin{pmatrix} \psi_{\alpha \ua}(n) \\ \psi_{\alpha \da} (n) \end{pmatrix}_{\beta \mu}. 
\end{equation}
It is useful to use compact notation and define a $4$$\times$$4$ wave-function matrix 
$[{\bm \Psi}(n)]$ whose elements are the spinor components $\psi_{\alpha \sigma,\beta \mu}(n)$ 
defined in Eq.~(\ref{Psi}), 
\begin{equation} \label{psi4x4}
[{\bm \Psi}(n)]_{\alpha \sigma,\beta \mu}=\psi_{\alpha \sigma,\beta \mu}(n),
\end{equation}
where the order of rows (counting from the top) or columns (counting from the left) is
 $(1 \ua, 1 \da,2 \ua, 2 \da)$.  

Now we define the $8$$\times$$8$ transfer matrices $T_n$, $n=-1,0,1,2,\ldots,N$, and a total transfer matrix $T$,
\begin{eqnarray} \label{Tn}
&& \begin{pmatrix} {\bm \Psi}(n) \\ {\bm \Psi}(n-1) \end{pmatrix}=T_{n-1} \begin{pmatrix}{\bm \Psi}(n-1) \\ {\bm \Psi}(n-2) \end{pmatrix}, \nonumber \\
 && T=T_NT_{N-1}[T_{N-2}....T_1]T_0T_{-1}  \mbox{ for } N>2,
\end{eqnarray}
while the product in the square parenthesis in Eq.~(\ref{Tn}) is the $8$$\times$$8$ identity matrix for $N=2$.  For an ordered lattice the spin-orbit (SO) potential is periodic and all the matrices forming the product inside the square parenthesis are identical, i.e.,  $T_1=T_2=\ldots =T_{N-2} \equiv \bar{T}$, and the product is equal to $\bar{T}^{N-2}$.  For the random ladder, all the matrices in the square parenthesis are different. The transfer matrices act on $8$$\times$$4$ wave function matrices.  The above construction implies that the total transfer matrix $T$ across the ladder satisfies
\begin{eqnarray} \label{transfer}
&& \begin{pmatrix} {\bm \Psi}(N+1) \\ {\bm \Psi}(N) \end{pmatrix} =T  \begin{pmatrix}{\bm \Psi}(-1) \\ {\bm \Psi}(-2) \end{pmatrix}.
\end{eqnarray}
Knowing the $8$$\times$$8$ transfer matrix $T$, one obtains the $4$$\times$$4$ transmission and reflection matrices $t$ and $r$ with elements $t_{ \alpha \sigma; \beta \mu}$ and $r_{ \alpha \sigma; \beta \mu}$. 

Starting from Eq.~(\ref{transfer}) we find, 
\begin{eqnarray} \label{psirt}
&& {\bm \Psi}(-1)=I_{4 \times 4}+r, \ \ {\bm \Psi}(-2)=e^{-i k}I_{4 \times 4}+e^{i k} r, 
\nonumber \\
&& {\bm \Psi}(N)=t, \ \ \  {\bm \Psi}(N+1)=e^{i k} t .
\end{eqnarray} 
This enables us to express $r$ and $t$ in terms of the four $4$$\times$$4$ blocks of $T$, denoted as $T_{ij}$, with $(i,j=1,2)$. The explicit expressions are:
\begin{eqnarray}
&& r=[e^{ik}(T_{21}-T_{12})+e^{2 i k}T_{22}-T_{11}]^{-1} \nonumber \\
&& [T_{11}+e^{-i k}T_{12}-e^{ik}T_{21}-T_{22}] \nonumber \\
&& t\mbox{=}e^{ik}T_{11}(I_{4 \times 4}\mbox{+}r)\mbox{+}e^{-ik}T_{12}(e^{-i k}I_{4 \times 4}\mbox{+}e^{ik}r).
\label{sol}
\end{eqnarray}
As a test of the correctness of these relations one can confirm the 
 time-reversal and unitarity constraints,
\begin{eqnarray} \label{unitarity1}
&& t'_{\alpha \nu \beta \mu}=(-1)^{\nu-\beta}t^*_{\beta \bar{\mu }\alpha \bar{\nu}}, 
\nonumber \\
&& r_{\alpha \nu \beta \mu}=(-1)^{\nu-\mu}r_{\beta \bar{\mu } \alpha \bar{\nu}}, \ r'_{\alpha \nu \beta \mu}=(-1)^{\nu-\mu}r'_{\beta \bar{\mu } \alpha \bar{\nu}}.
\nonumber \\
&& {\mathrm{Tr}}[t^\dagger t+r^\dagger r]=4, \quad
\end{eqnarray}
Here $\bar {\sigma}=-\sigma$, and $t'$ and $r'$ are the transmission and reflection matrices for scattering of incoming electrons from the right.  Note that these relations connect matrix elements of the transmission matrices {\em on different sides of the sample}, and matrix elements of the reflection matrices {\em on the same side of the sample}.  These relations imply the absence of spin-flip in the reflection amplitude of the same channel, i.e., $r_{\alpha \nu \alpha \bar {\nu}}=0$, and also that the diagonal elements of the reflection matrix are equal, $r_{\alpha \ua, \alpha \ua}=r_{\alpha \da, \alpha \da}$ for each channel.  We shall see below that these relations also imply the absence of transmitted and reflected polarizations. 

It remains to determine the $8$$\times$$8$ local transfer matrices $\{ T_n \}$. A glance at Fig.~1 suggests that there are three kinds of sites: (1) For $n=-1$ and $n=N$ the coordination number is 2 and the two links attached to it are bare. (2) For $n=0$ and $n=N-1$ the coordination number is 3, one link is bare and two links are ``dressed'' with SU(2) hopping matrices. (3) For $n=1,2,\ldots, N-2$ the coordination number is 3, and all three links are ``dressed'' with SU(2) hopping matrices. This respectively requires three slightly different definitions. For this purpose it is useful to define the following $4$$\times$$4$ matrices:
$$ 
X \equiv \begin{pmatrix} 0&e^{i \lambda \sigma_x} \\ e^{-i \lambda \sigma_x}&0 \end{pmatrix}, \ \ Z \equiv I_{2 \times 2} \otimes e^{i \lambda \sigma_z}. 
$$
After some algebra we find,
\begin{eqnarray} \label{Tnexplicit}
&& T_{-1}=\begin{pmatrix} -\tfrac{\varepsilon}{t}&-1\\
1& 0 \end{pmatrix}, \quad  T_0=\begin{pmatrix} -Z^\dagger(\tfrac{\varepsilon}{t}+X)&-Z\\
1& 0 \end{pmatrix}, \nonumber \\
&& T_n=\begin{pmatrix} -Z^\dagger(\tfrac{\varepsilon}{t}+X)&-Z^2\\
1& 0 \end{pmatrix}, \ n=1,2,\ldots,N-2, \ \nonumber \\
&& T_{N-1}=\begin{pmatrix} -(\tfrac{\varepsilon}{t}+X)&-Z\\
1& 0 \end{pmatrix}, \ \ T_N=T_{-1}~,
\end{eqnarray}
where every entry in these matrices is a $4$$\times$$4$ matrix in channel$\otimes$spin space.

\section{Aharonov--Casher Phase for a square subject to a homogeneous perpendicular elecric field} \label{AC_phase}
\begin{figure} 
\centering{\includegraphics[width=0.45\textwidth]{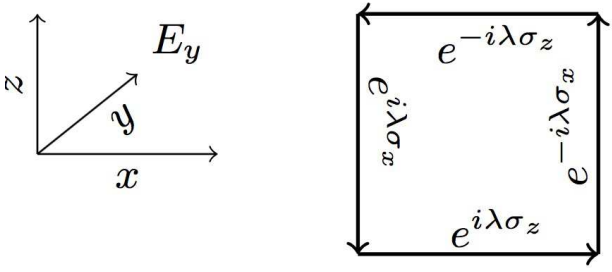}}
\caption{Electron hopping on the corners of a square in the $x$-$z$ plane subject to a homogeneous electric field $E\hat {\bf y}$.  Rashba (actually Pauli) SO coupling generates SU(2) hopping matrix elements $e^{\pm  i \beta \sigma_z}$ along the horizontal links and $e^{\mp  i \beta \sigma_x}$ along the vertical links.}
\label{SquareE}
\end{figure}

\begin{figure}
\centering
\centering{\includegraphics[width=0.35\textwidth]{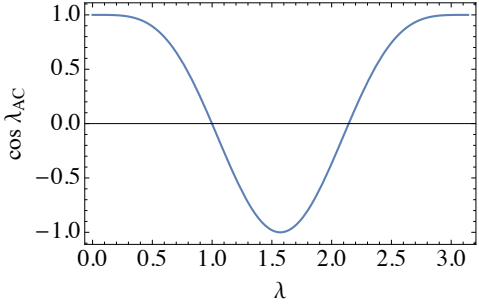}}
\caption{(Color online) $\cos \lambda_{\mathrm{AC}}$ versus the SO coupling strength $\lambda$ for the square system  in Fig.~\ref{SquareE} [see Eq.~(\ref{flux9})].}
\label{ACphase}
\end{figure}

Consider a closed contour in the form of a square of side $a=1$ lying on the $x$-$z$ plane subject to a constant electric field $E$ along $y$. Within a tight-binding model an electron hops between its corners, as in Fig.~\ref{SquareE}. Upon traversing  the square counterclockwise, starting and ending at (0,0), the wave function of an electron gains an SU(2) phase factor 
\begin{equation} \label{flux8} 
F \equiv e^{i \lambda \sigma_z}e^{-i \lambda \sigma_x}e^{-i \lambda \sigma_z}e^{-i \lambda \sigma_x} ,
\end{equation}
where $\lambda=eEa/(4mc^2)$ is dimensionless. Since $F$ is an SU(2) matrix, it can be written as
\begin{equation} \label{F}
F=e^{i \lambda_{\mathrm {AC}} \, \hat{\bf n} \cdot {\bm \sigma}} = \cos \lambda_{\mathrm {AC}}+i \sin \lambda_{\mathrm {AC}} \, \hat{\bf n} \cdot {\bm \sigma},
\end{equation}
where $\lambda_{\mathrm {AC}}$ is the Aharonov--Casher phase.  The value of $\lambda_{\mathrm {AC}}$ is given by, 
\begin{equation} \label{flux9}
\cos \lambda_{\mathrm {AC}} = \tfrac{1}{2} \mbox{Tr}F = 1-2 \sin^4 \lambda .
\end{equation}
The fact that calculation of $\lambda_{\mathrm {AC}}$ involves a trace indicates that this expression is gauge invariant. Note that the relation between $\lambda$ and $\lambda_{\mathrm {AC}}$ is highly non-linear, so that the definition of SU(2) flux seems problematic.  In Fig.~\ref{ACphase} $\cos \lambda_{\mathrm{AC}}$ is plotted as a function of the SO coupling strength $\lambda$ (which is identical on all four links).

\section{Absence of polarization in systems with two 1D leads}  \label{Sed:NoP}
Following Eq.~(\ref{G3}), we pointed out that if a sample is composed of strictly 1D helix with SO coupling, it is not possible to obtain a non-vanishing polarization because in the absence of a closed loop, the SU(2) vector potential can be eliminated by a gauge transformation. This statement is based on the formulation of the SO coupling within an SU(2) gauge formalism, namely, it relies on an approximation. Here we show that due to unitarity and time reversal invariance, this statement is exact.  Moreover, it holds for any system with a maximum of two strictly 1D leads, including systems with closed loops such as in a ring interferometer.  To show this, let us denote by $t,r$ ($t',r'$) the 2$\times$2 transmission and reflection matrices for an electron scattering impinging on the sample from lead 1 (2). Then, the 4$\times$4 $S$ matrix and its unitarity relation read,
\begin{equation}\label{Sunit}
S=\begin{pmatrix} r&t'\\t&r' \end{pmatrix}, \ \ SS^\dagger=\begin{pmatrix} r&t'\\t&r' \end{pmatrix}\begin{pmatrix} r^\dagger &t^\dagger \\ t'^\dagger &r'^\dagger \end{pmatrix}={\bf 1}_{4 \times 4}. 
\end{equation} 
Let us focus our attention on a non-diagonal element of the unitarity relation (a 2$\times$2 matrix),
\begin{equation} \label{unit12}
[SS^\dagger]_{12}= r t^\dagger+t'r'^\dagger={\bf 0}_{2 \otimes 2}.
\end{equation}
The constraints (\ref{unitarity1}) imposed on the transmission and reflection matrices show that,
\begin{eqnarray} \label{constraintstr}
 && r=\rho e^{i \theta} {\bf 1}_{2 \otimes 2}, \ \ r'=\rho e^{i \theta'} {\bf 1}_{2 \otimes 2}, \ \ 
 t=\begin{pmatrix} \tau e^{i \alpha}&\eta e^{i \beta}\\ \eta e^{i \gamma} & \tau e^{i \delta} \end{pmatrix}, \nonumber \\
 && \rho, \tau, \eta >0, \ \ \rho^2+\tau^2+\eta^2=1, \nonumber \\
 && (\beta-\alpha)-(\gamma-\delta)=(2n+1)\pi \ \ , n=\mbox{integer}.
\end{eqnarray}
First, it is easy to see that the reflected polarization vanishes because $r$ is proportional to the unit matrix $I_{2 \times 2}$ and hence 
Tr$[r \, {\bm \sigma} \, r^\dagger]=0$. As for the transmitted polarization, we use Eq.~(\ref{unitarity1}) and express the elements of $t'$ in terms of the element of $t$, and then use Eq.~(\ref{unit12}) to obtain
the third equation of (\ref{constraintstr}). For an {\it unpolarized incoming beam of electrons}, this implies the vanishing of the transmitted polarization, i.e., 
\begin{eqnarray}
&& g {\bf P}_{T,x}=\mbox{Tr}[t \sigma_x t^\dagger] = 2 \tau \eta [\cos(\alpha-\beta)+\cos(\gamma-\delta)]=0, \nonumber \\
&& g {\bf P}_{T,y}=\mbox{Tr}[t \sigma_y t^\dagger] = 2 \tau \eta [\sin(\alpha-\beta)+\sin(\gamma-\delta)]=0, \nonumber \\
&& g {\bf P}_{T,z}=\mbox{Tr}[t \sigma_z t^\dagger] =0, \ \mbox{independent of angles}.
\label{Polxyz}
\end{eqnarray}

\noindent
\underline{Numerical example}\\
\begin{figure}[htb]
\centering{\includegraphics[width=0.45\textwidth]{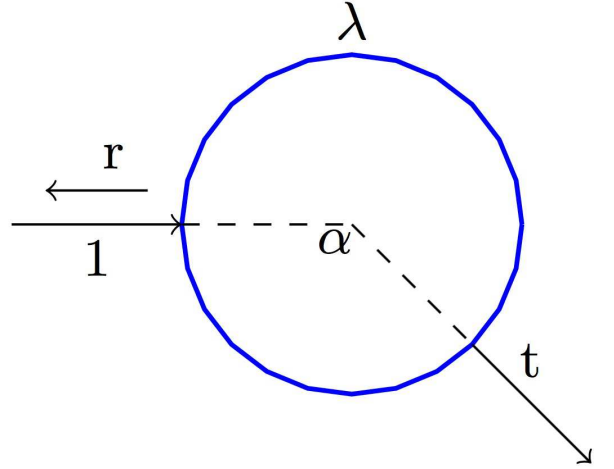}}
\caption{(Color online) 1D ring interferometer (schematic). Electrons approaching the sample from left at polar angle $\theta=0$ are partially reflected and partially transmitted at the second lead at polar angle $\theta=\alpha$ (reflection matrix $r$ and transmission matrix $t$).  The ring is subject to a homogeneous perpendicular electric field, so that the Rashba SO mechanism generates an effective magnetic field along the radial direction $\hat {\bf n}(\theta)=(\cos \theta, \sin \theta,0)$.  The local phase factor for polar angle $\theta$ is $e^{i \lambda \, \hat {\bf n}(\theta) \cdot {\bm \sigma}}$.}
\label{ring_int}
\end{figure}
As a numerical example, consider a ring interferometer of radius $R=1$ with two 1D leads 
(source lead at angle $\theta=0$ and drain lead at $\theta=\alpha=1.73$),
subject to a perpendicular electric field leading to Rashba SO coupling, as shown in
Fig.~\ref{ring_int}. 
The SO strength is taken to be $\lambda=2.3$ and 
the wave number is $k=1.25$.  The solution of the 
scattering problem yields the transmission and reflection matrices. 
$$ t=\begin{pmatrix} 0.308365+0.0888927i & -0.253756 + 0.791537i\\
 0.267907 + 0.786861i & 0.306722 - 0.0944069i \end{pmatrix} , $$
$$ r = (0.00406962 + 0.453948i)\begin{pmatrix}
 1 & 0 \\
 0 & 1 \end{pmatrix},
$$ 
The dimensionless conductance  is $g$=Tr$[t^\dagger t]$=1.58783. 
Using Eq.(\ref{Polxyz}) the reader can easily verify that all the transmitted 
(and reflected) polarizations vanish.

\section{Spin Density, Current and Torque}  \label{App_spin-torque}

Spin density, spin current and spin-torque are three central quantities in contemporary spintronics. In this section we show that our ladder model is an appropriate theoretical framework for studying these quantities. As a byproduct, we derive a useful relation between the transmitted and reflected polarization vectors and the volume integrated spin-torque. 

\noindent 
 \underline{The spin density} at site $(\alpha,n)$ resulting from an incoming wave in state $|\beta \mu \ra$ is a {\em local} vector field in spin space defined as \cite{Qian-Niu,band-avishai-2}, ${\bf S}_{\alpha; \beta \mu}(n) =  \psi^\dagger_{\alpha; \beta \mu}(n){\bf s} \psi_{\alpha; \beta \mu}(n)$.  Here ${\bf s}$ is the electron spin operator in the Heisenberg representation and the spinor $\psi_{\alpha; \beta \mu}(n)$ is defined in Eq.~(\ref{Psi}).  In the leads ($n<0$ for the left lead and $n \ge N$ for the right lead), where SO coupling is absent,  ${\bf S}_{\alpha; \beta \mu}(n)={\bf S}_{\alpha; \beta \mu}$ is independent of $n$, and the spin densities on the left and right are respectively expressible  in terms of the reflection and transmission amplitudes:
\begin{equation} \label{SLR}
{\bf S}_{\alpha; \beta \mu}(n)=\begin{cases}  \mu \delta_{\alpha \beta}+ \sum_{\sigma \sigma'}r_{\alpha \sigma \beta \mu}^*[{\bf s}]_{\sigma \sigma'} r_{\alpha \sigma' \beta \mu} \ \ (n<0) \\
\sum_{\sigma \sigma'}t_{\alpha \sigma \beta \mu}^*[{\bf s}]_{\sigma \sigma'} t_{\alpha \sigma' \beta \mu} \ (n \ge N)
\end{cases}.
\end{equation}

\noindent
\underline{The spin current} on chain $\alpha$ and site $n$ resulting from an incoming wave in state $|\beta \mu \ra$ is a local quantity (a tensor field with  Cartesian and spin components) defined as \cite{Qian-Niu, band-avishai-2} 
$\mathbb{J}_{\alpha; \beta \mu}(n)= \tfrac{1}{2} {\mathrm{Re}} [\psi^\dagger_{\alpha; \beta \mu}(n) \{ {\bf s}, {\bf v} \}  \psi_{\alpha; \beta \mu}(n)] $ (where ${\bf v}$ is the velocity operator). For $n<0$ and $n \ge N$ we have,
\begin{equation} \label{bfJ}
\mathbb{J}_{\alpha; \beta \mu}(n) \! \! = \! \!
\begin{cases} \sin k \, \{ \mu \delta_{\alpha \beta} - \sum_{\sigma \sigma'}r_{\alpha \sigma \beta \mu}^*[{\bf s}]_{\sigma \sigma'} r_{\alpha \sigma' \beta \mu}  \} \ \ (n<0) \\
 \sin k \sum_{\sigma \sigma'}t_{\alpha \sigma \beta \mu}^*[{\bf s}]_{\sigma \sigma'} t_{\alpha \sigma' \beta \mu} \ (n \ge N),
\end{cases}
\end{equation} 
hence $\mathbb{J}_{\alpha; \beta \mu}(n)$ is independent of $n$. The {\it total spin current in the leads} (where SO is absent) is obtained after summing on initial conditions $(\beta \mu)$ 
and adding the contributions from both chains:
\begin{equation} \label{Jtot}
\mathbb{J} = \begin{cases}  \sin k \, {\mathrm{Re}} \{ {\mathrm{Tr}} [t^\dagger {\boldsymbol \Sigma} t] \} \equiv \mathbb{J}^T(n)  \ \ \ (n \ge N) \\ 
 -  \sin k \, {\mathrm{Re}} \{ {\mathrm{Tr}} [r^\dagger{\boldsymbol \Sigma} r] \} \ \equiv \mathbb{J}^R(n)  \ \ \ (n<0).  
 \end{cases}
\end{equation}
Here ${\boldsymbol \Sigma}=\tfrac{1}{2} I_{2 \times 2} \otimes {\bm \sigma}$ 
(defined after Eq.~(\ref{BAPol})), and $\mathbb{J}^T$ and $\mathbb{J}^R$ are the transmitted and reflected spin currents. In the present case the spin current tensor field has only a single Cartesian component, namely, it is directed along $x$ 
(see Fig.~\ref{LadderE}), and has spin components according to ${\boldsymbol \Sigma}$.  
We shall sometimes denote $\mathbb{J}_i = \mathbb{{\bf J}}_i$ in order to remind us that the $i$th component of the spin current in Euclidean space is still a vector in spin space. Following Eq.~(\ref{Jtot}), for $n>N-1$ we have $\mathbb{{\bf J}}_x=\mathbb{\bf J}_x^T$ 
(the transmitted spin current) while for $n<0$ we have $\mathbb{{\bf J}}_x=\mathbb{\bf J}_x^R$ 
(the reflected spin current).  The sign of the components of the spin current depends on 
its directions in Euclidean {\it and} in spin spaces. Explicitly, let us denote
the $z$ component (in spin space) of $\mathbb{\bf J}_x^T$ and $\mathbb{\bf J}_x^R$
   by $\mathbb{J}_x^{zT}$ and $\mathbb{J}_x^{zT}$.  Then we have
\begin{equation} \label{decompose}
 \mathbb{ J}_x^{zT}= \mathbb{ J}_x^{zT\ua}+\mathbb{ J}_x^{zT\da}, 
 \quad  \mathbb{J}_x^{zR}=\mathbb{J}_x^{zR\ua}+\mathbb{ J}_x^{zR\da},
\end{equation}
and note that
\begin{equation} \label{Jupdown}
  \mathbb{ J}_x^{zT \ua}>0, \ \ \mathbb{ J}_x^{zT\da}<0,  \quad \mathbb{ J}_x^{zR \ua}<0, \ \ \mathbb{ J}_x^{zT\da}>0.
\end{equation}

\noindent
\underline{The spin-torque} is a vector in spin space:
\begin{eqnarray} \label{spin-torque}
&& {\bm \tau}_{\alpha; \beta \mu}(n)= {\mathrm{Re}} \{\psi^\dagger_{\alpha; \beta \mu}(n) \frac {d {\bf s}}{dt}  \psi_{\alpha; \beta \mu}(n) \} \nonumber \\
&& = {\mathrm{Re}} \{ \frac{1}{i \hbar}\psi^\dagger_{\alpha; \beta \mu}(n)  [{\bf s},H] \psi_{\alpha; \beta \mu}(n) \}. 
\end{eqnarray}
 The experimentally relevant observable spin-torque is obtained after summation over all initial conditions, i.e., ${\bm \tau}_\alpha(n) = \sum_{\beta \mu} {\bm \tau}_{\alpha; \beta \mu}(n)$.  Outside the SO interaction region $\{ n<0 \} \cap \{n \ge N \}$, $[{\bf s},H] = 0$, and ${\bm \tau}_\alpha(n)$ vanishes.  Generically, in the region, $0 \le n \le N-1 \}$ where SO is active, ${\bm \tau}_\alpha(n) \ne 0$ because $[{\bf s},H] \ne 0$ {\it even in the static case}.  Of special interest is the volume integrated spin-torque (see Ref.~\cite{Qian-Niu}) which, for the discrete tight-binding geometry, takes the form ${\bf T} = \sum_{\alpha n} {\bm \tau}_\alpha(n)$.  
  \ \\
  \ \\
\noindent
\underline{Relation between spin flux and spin-torque.}
Upon using the continuity equation for the spin current in the stationary case (where $\partial {\bf S}_{\alpha;\beta \mu} (n)/\partial t = 0$) \cite{Qian-Niu}, and summing over initial conditions $(\beta \mu)$, we obtain
\begin{eqnarray} 
\label{stat-continuity}
&& {\bm \nabla} \cdot  \mathbb{J}_\alpha (n) = - {\bm \tau}_\alpha(n), \nonumber \\ 
&& [{\mbox{in the continuum limit, }} {\bm \nabla} \cdot  \mathbb{J}(\bfr) = - {\bm \tau}(\bfr)],
\end{eqnarray}
which is valid anywhere (i.e., for any $n$, including inside the sample $0 \le n \le N-1$ where SO is active).
Summing over $(\alpha,n)$ in Eq.~(\ref{stat-continuity}) we obtain\cite{Hyman},
\begin{equation} \label{div}
\sum_{\alpha,n} {\bm \nabla} \cdot  \mathbb{J}_\alpha (n)=-{\bf T},
\end{equation}
 Using the divergence theorem,  the LHS of Eq.~(\ref{div}) equals the total spin flux ${\bm \Phi}$ out of the sample into the left and right leads. Note that ${\bf \Phi}$ is a vector in spin space.  In Ref.~\cite{Qian-Niu} it is noted that under certain symmetry constraints, ${\bf T}=0$.  In our tight-binding model, the spin currents for $n \ge N$ and for $n<0$ are perpendicular to the links and there is no spin current along $z$ (through the chains).   Therefore, the transmitted and reflected spin fluxes through the sample borders are ${\bm \Phi}^T=a \mathbb{\bf J}_x^T$ and ${\bm \Phi}^R=a \mathbb{\bf J}_x^R$ where $a=1$ is the length of the links.  Then, if ${\bf T}=0$,  the  vector field $\mathbb{\bf J}^{x}(\bfr)$  (directed along $x$, and having spin component $x,y,z$) are flux-less; the same spin current that enters the sample on one side leaves it on the other side and the total spin flux vanishes.  In the generic case, however,  where ${\bf T} \ne 0$, the total spin flux is related the transmitted and reflected currents and to the volume-integrated spin-torque as,
\begin{equation} \label{TorqueFlux}
{\bm \Phi} = a (\mathbb{\bf J}_x^T-\mathbb{\bf J}_x^R)= -{\bf T}.
\end{equation}
Equation~(\ref{TorqueFlux}) gives us a direct and useful relation between the measurable current {\it outside the interaction region (the sample)}, and the wave functions {\it inside the sample} which appears in the calculations of the spin-torque, Eq.~(\ref{spin-torque}).  In addition, our calculations on similar systems indicate that ${\bf T}$ is correlated also with the charge conductance $g$, i.e., the charge conductance reflects the underlying spin physics.   Although Eq.~(\ref{TorqueFlux}) is a direct consequence of the continuity conditions, we have not seen its derivation elsewhere.


\begin{thebibliography}{99}
\bibitem{Ozols_mod} In Eq.~(21) of Ref.~\cite{Ozols}, it is suggested to take $\sin \theta= \sqrt{\xi}$ where $\xi \in [0,1]$ is a random number, from which it follows that $\cos \theta=\pm (1-\xi)$. However, the choice of the sign is not specified therein, so we choose a random sign for it.
\end{thebibliography}
\end{document}